\documentclass[%
aip,
sd,%
amsmath,amssymb,
reprint,%
onecolumn,%
]{revtex4-1}

\usepackage{graphicx}% Include figure files
\usepackage{dcolumn}% Align table columns on decimal point
\usepackage{bm}% bold math
\usepackage{ulem}
\usepackage{color}
\newsavebox{\astrutbox}
\sbox{\astrutbox}{\rule[-5pt]{0pt}{20pt}}

\begin{document}

\preprint{AIP/123-QED}

\title[Nonlinear internal wave penetration via parametric subharmonic instability]{Nonlinear internal wave penetration via parametric subharmonic instability}% Force line breaks with \\
%\thanks{Footnote to title of article.}

\author{S. J. Ghaemsaidi}
  \email{sjsaidi [at] mit [dot] edu}
 \affiliation{Department of Mechanical Engineering, Massachusetts Institute of Technology, Cambridge, MA 02139, USA.}%Lines break automatically or can be forced with \\
\author{S. Joubaud}
\affiliation{Laboratoire de Physique, {E}cole Normale Sup\'{e}rieure de Lyon, Universit\'{e} de Lyon, CNRS, 46 All\'{e}e d'Italie, F-69364 Lyon cedex 07, France
}%
\author{T. Dauxois}%
\affiliation{Laboratoire de Physique, {E}cole Normale Sup\'{e}rieure de Lyon, Universit\'{e} de Lyon, CNRS, 46 All\'{e}e d'Italie, F-69364 Lyon cedex 07, France
}%
\author{P. Odier}
\affiliation{Laboratoire de Physique, {E}cole Normale Sup\'{e}rieure de Lyon, Universit\'{e} de Lyon, CNRS, 46 All\'{e}e d'Italie, F-69364 Lyon cedex 07, France
}%cedex
\author{T. Peacock}
 \affiliation{Department of Mechanical Engineering, Massachusetts Institute of Technology, Cambridge, MA 02139, USA.}

\date{\today}% It is always \today, today,

\begin{abstract}

We present the results of a laboratory experimental study of an internal wave field generated by harmonic, spatially-periodic boundary forcing from above of a density stratification comprising a strongly-stratified, thin upper layer sitting atop a weakly-stratified, deep lower layer. In linear regimes, the energy flux associated with relatively high frequency internal waves excited in the upper layer is prevented from entering  the lower layer by virtue of evanescent decay of the wave field. In the experiments, however, we find that the development of parametric subharmonic instability (PSI) in the upper layer transfers energy from the forced primary wave into a pair of subharmonic daughter waves, each capable of penetrating the weakly-stratified lower layer. We  find that around $10\%$ of the primary wave energy flux penetrates into the lower layer via this nonlinear wave-wave interaction for the regime we study.

\end{abstract}

\pacs{}
\keywords{internal waves, parametric subharmonic instability, wave-wave interactions, stratified fluids}
                              
\maketitle

% Paragraph 1: internal waves & system description
Internal waves are prevalent in the oceans by virtue of gravitationally-stable background temperature and salinity gradients. Typically, the oceanic water column is characterized by a mixed layer of uniform density overlying a strongly-stratified near-surface layer, beneath which lies the weakly stratified deep ocean. Of  substantial interest, and a motivation for this study, is how internal waves excited by motions in the mixed layer may penetrate into the deep ocean, where they may potentially drive deep-ocean mixing \citep{MunkWunsch1998,Wunsch2004}. 

% Paragraph 2: traditional linear mechanisms
There have been many studies regarding the propagation of linear internal wave fields through nonuniform density stratifications, as characterized by the vertical variation of the local buoyancy frequency  $N(z) = \sqrt{\left(-g/\rho\right) \mathrm{d} \rho/\mathrm{d}z}$, where $z$ is the coordinate aligned antiparallel with gravity $g$ and $\rho(z)$ is the background density profile. A standard procedure that provides substantial insight is to calculate the transmission properties through piecewise linear density gradients by matching boundary conditions at interfaces\cite{Sutherland2004}. A notable advance on such methods was the development of an analytical approach to tackle propagation through continuous stratifications without requiring any assumptions regarding the relative scale of the waves and variations of the background stratification \cite{Nault2007}. This approach was subsequently extended to address the propagation of internal wave beams through nonuniform stratifications \cite{Mathur2009}, the predictions being validated by close agreement with laboratory experiments. All the aforementioned examples being linear models, however, the propagating wave field does not experience wave-wave interactions. 

% Paragraph 3: what is PSI / previous studies
It is well established that nonlinearities can give rise to wave-wave interactions and resonant growth via the so called parametric subharmonic instability (PSI) \citep{Mied1976,Drazin1977,Koudella2006}. This process takes internal wave energy from a primary, `mother' wave of frequency $\omega_0$ and wavenumber $\boldsymbol{k}_0$, and distributes it among a pair of `daughter' waves of frequencies $\omega_1$ and $\omega_2$, and wavenumbers $\boldsymbol{k}_1$ and $\boldsymbol{k}_2$, with the requirements that  $\omega_1+\omega_2=\omega_0$ and $\boldsymbol{k}_1+\boldsymbol{k}_2=\boldsymbol{k}_0$. The subharmonic daughter waves birthed by PSI are characterized by smaller vertical length scales, and thus higher vertical shear, and slower group velocities. In light of this, field studies \citep{Alford2007,MacKinnon2013} and numerical simulations \citep{MacKinnon2005,Gayen2013} have sought to investigate PSI driven processes to explore its potential to drive mixing processes in the ocean. Complementary laboratory studies in uniform stratifications have  demonstrated conditions under which internal wave fields become unstable to minute perturbations via PSI\citep{Joubaud2012,Bourget2013}.

% Paragraph 4: what we are studying
In this letter, we present the results of a laboratory experiment in which internal waves are excited in a relatively shallow, upper layer of high buoyancy frequency, $N_1$, sitting atop a deeper layer of low buoyancy frequency, $N_2$. The internal wave field is excited at a relatively high frequency $\omega_0$ (i.e. $N_2 < \omega_0 < N_1$) via boundary forcing imposed at the top of the upper layer; as such, the wave field excited at the forcing frequency is confined to the upper layer because it is evanescent in the lower layer. By observing the evolving wave dynamics over sufficiently long time scales, we study the development and consequences of PSI in this setting.

%%%%%%%%%%%%%%%%%%%%%%%%%%%%%%%%%%%%%%%%%%%%%%%%%%%%%%%%%
%%%%%%%%%%%%%%%%%%%%%%%%%%%%%%%%%%%%%%%%%%%%%%%%%%%%%%%%%

\begin{figure}
\centering
\includegraphics[width=1\columnwidth]{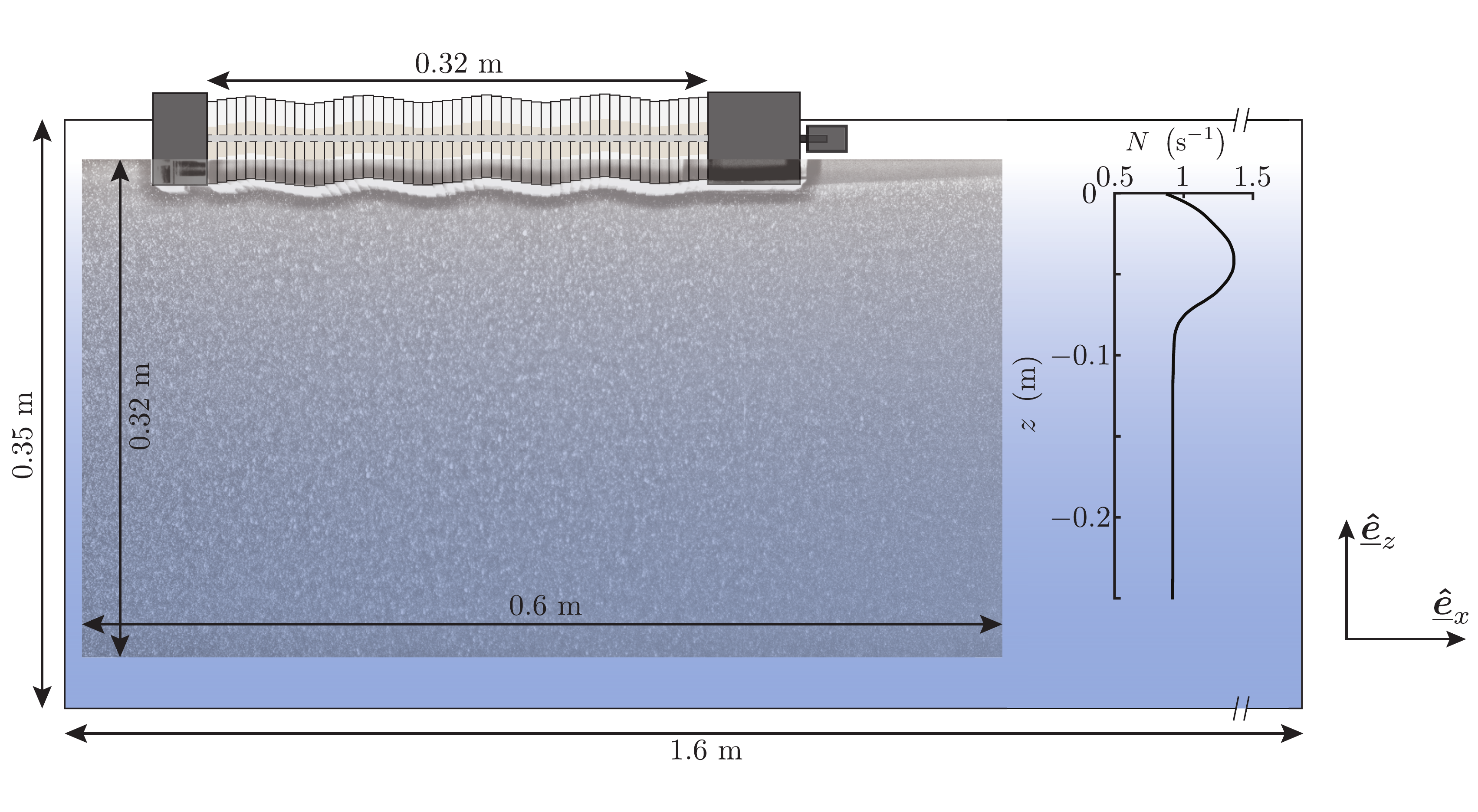}
\caption{Schematic of the experimental configuration. A sample particle image is overlaid to show the field of view captured by the CCD camera. The measured buoyancy frequency profile $N(z)$ is overlaid to the right.}
\label{fig:experiment}
\end{figure}

% Paragraph 5
A schematic of the experimental configuration, incorporating a plot of the measured experimental stratification, is presented in figure \ref{fig:experiment}. A $1.6~\mathrm{m} \times 0.4~\mathrm{m} \times 0.17~\mathrm{m}$ acrylic tank was filled with salt-stratified water of depth $0.35~\mathrm{m}$ using the double bucket method. The density stratification, measured using a conductivity-temperature probe, comprised a roughly $8~\mathrm{cm}$ thick upper layer of peak buoyancy frequency $N_1 \approx 1.25~\mathrm{s}^{-1}$ and an approximately $25~\mathrm{cm}$ deep lower layer of buoyancy frequency $N_2 \approx 0.93~\mathrm{s}^{-1}$. The nonuniform stratification was established by first filling the top and subsequently the bottom stratified layer, adjusting the bucket density ratios in between these two filling processes. 

% Paragraph 6
A horizontal wave generator was mounted atop the tank in such a manner that the vertical motion of the array of plates within the generator forced the upper few millimetres of the stratification; the wave generator had a rightward phase velocity and thereby produced only left-to-right, downward propagating internal waves. The vertical end walls were sufficiently far away that reflections did not affect the field of view for the time scales considered. The wave generator spanned $32~\mathrm{cm}$ and contained four horizontal wavelengths of $\lambda = 7.7~\mathrm{cm}$ in addition to several buffer plates that smoothly tapered from the amplitude of forcing to zero at either end of the generator.  A traveling wave was excited by the wave generator, imposing a vertical velocity on the stratification of the form $w(x,z=0,t) = A \omega_0 \cos \left(k_0 x - \omega_0 t\right)$, where $x$ is the horizontal coordinate, $t$ is time, the amplitude $A = 2.5~\mathrm{mm}$, the frequency $\omega_0\approx1.03 ~\mathrm{s}^{-1}$ and the horizontal wavenumber $k_0 = 2\pi/\lambda \approx 81.6~\mathrm{m}^{-1}$.  By choosing the forcing frequency of the wave generator such that $N_2 < \omega_0 < N_1$, the upper layer acted as a waveguide supporting propagating internal waves, with evanescent decay taking place in the lower $N_2$ layer. The characteristic Reynolds number for the experiment was $\mathrm{Re} = A\omega_0/\nu k_0 \approx 30$, where the kinematic viscosity is $\nu \approx 1 \times 10^{-6}~\mathrm{m}^2~\mathrm{s}^{-1}$; we thus expect viscosity to be present and have a modest, not dominant, role in the experiments.

% Paragraph 7
Particle image velocimetry (PIV) was used to obtain velocity field data. Prior to filling, the water was seeded with hollow glass spheres of mean diameter $10~\mu\mathrm{m}$, which were subsequently illuminated by an approximately $2~\mathrm{mm}$ thick laser sheet. The evolving motion of the internal wave field was recorded by a CCD camera positioned normal to, and $1.5~\mathrm{m}$ away from, the front face of the tank. The measurement field of view covered a $60~\mathrm{cm} \times 32~\mathrm{cm}$ (2400 pixels $\times$ 1300 pixels) area below the wave generator (as illustrated in figure \ref{fig:experiment}). The nominally two-dimensional velocity fields were visualized in the mid-plane of the tank, so as to minimize sidewall effects. Images were recorded at a frame rate corresponding to 32 images per oscillatory period. These images were analyzed using the LaVision DaVis 7.2 PIV processing software package to obtain the experimental velocity fields.  

%%%%%%%%%%%%%%%%%%%%%%%%%%%%%%%%%%%%%%%%%%%%%%%%%%%%%%%%%
%%%%%%%%%%%%%%%%%%%%%%%%%%%%%%%%%%%%%%%%%%%%%%%%%%%%%%%%%

\begin{figure}
\centering
\includegraphics[width=1\columnwidth]{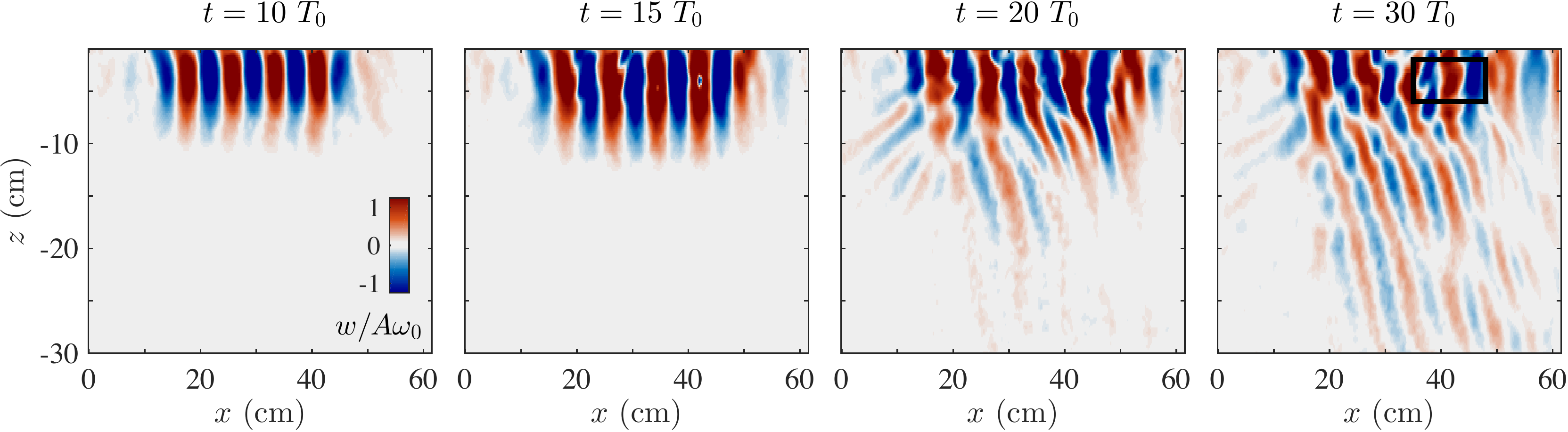}
\caption{Snapshots of the vertical velocity wave field, normalized by the characteristic velocity scale $A \omega_0$, at successive times illustrating the evolution of the wave field due to the development of PSI in the upper $N_1$ layer. The rectangular boundary drawn in the rightmost panel denotes the region used to perform the time-frequency analysis presented in figure \ref{fig:spectrogram}. The coordinates of the boundary are $x \in [35,48]$~cm and $z \in [-6,-2]$~cm; the box was chosen to be sufficiently large to capture an extended vertical and horizontal section of the pycnocline, so as to sample all the wave components.}
\label{fig:time evolution}
\end{figure}

% Paragraph 8
Snapshots illustrating the temporal evolution of the experimental wave field are presented in figure \ref{fig:time evolution}. As demonstrated by the snapshots, steady state is established by $\sim 10$ forcing periods, after which an additional $\sim 5-10$ forcing periods are needed for PSI to be clearly manifest, but this is strongly dependent on the amplitude of the primary wave. At time  $t = 10T_0$, where $T_0=2\pi/\omega_0$, waves are initially restricted to the upper $N_1$ layer and experience evanescent decay below, consistent with linear expectations. At $t = 15T_0$, the emergence of PSI becomes evident in the $N_1$ layer in the form of a modest wrinkling of the primary wave field, these perturbations being the subharmonics superposed onto the primary wave field. PSI is clearly established by $t = 20T_0$, by which time the downward propagating daughter waves penetrate the lower $N_2$ layer at two different propagation angles, one to the right and one to the left, thus revealing the presence of two daughter waves oscillating at a frequency less than $N_2$. By $t = 30T_0$, the daughter waves are well established throughout the lower $N_2$ layer, with PSI continuing to operate in the upper $N_1$ layer.

\begin{figure}
\centering
\includegraphics[width=0.7\columnwidth]{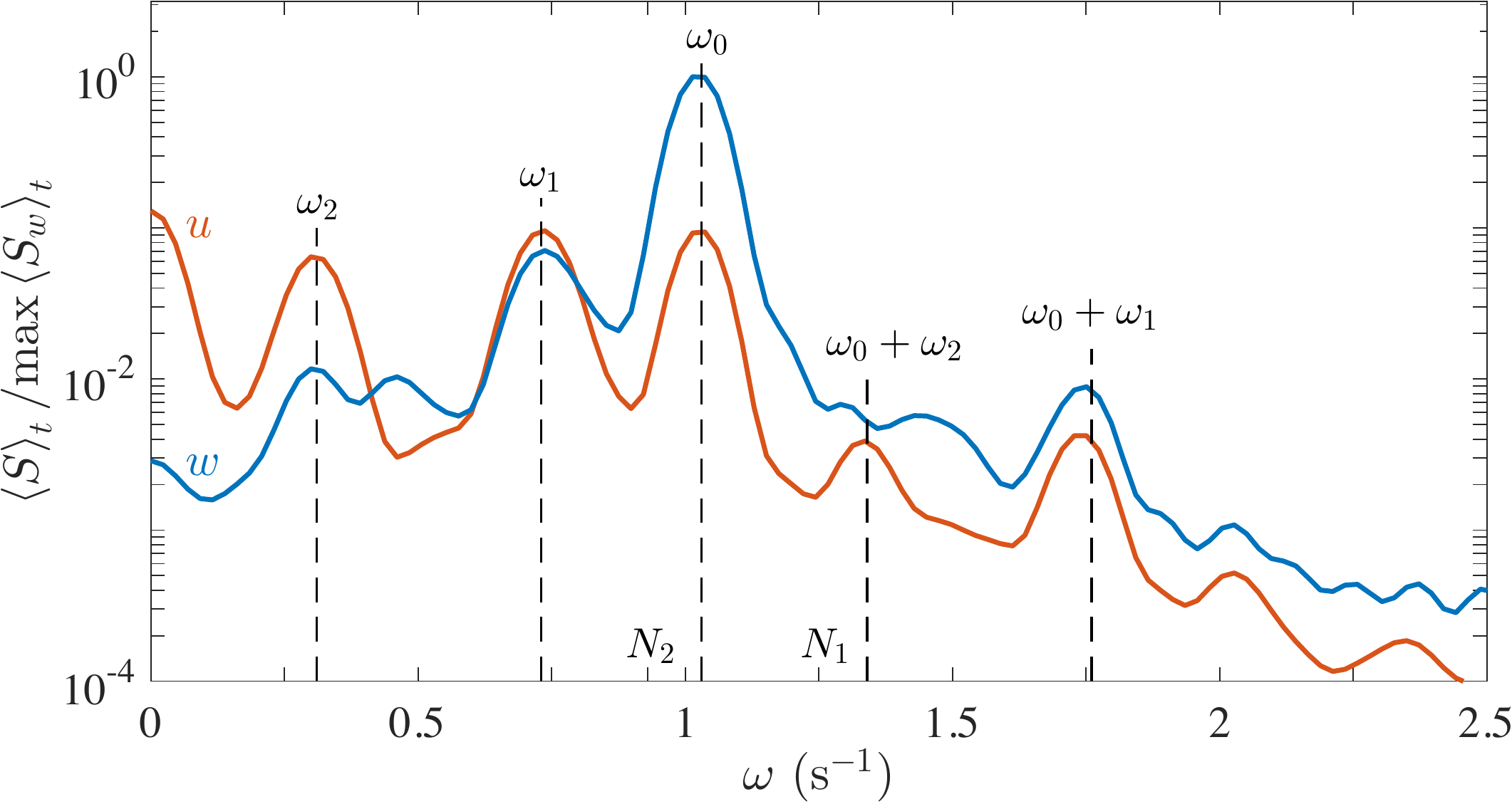}
\caption{Time-averaged spectrograms, $S\left(\omega,t\right)$, for the $u$ (red) and $w$ (blue) velocity components normalized by the maximum quantity.}
\label{fig:spectrogram}
\end{figure}

% Paragraph 9
To shed light on the frequency content of the wave field, we calculate and average the time-frequency spectra \citep{Flandrin1999} of the wave field in the rectangular domain indicated in the rightmost panel of figure \ref{fig:time evolution}; this region is chosen because it contains all wave components that arise throughout the duration of the experiment. More specifically, spectrograms are computed for both the $u$ and $w$ velocity components by time-averaging the last five periods of the data set, by which time PSI had clearly been established. Figure \ref{fig:spectrogram} presents the  results of the analysis, with the spectra having prominent peaks at the forcing frequency, $\omega_0 \approx 1.03~\mathrm{s}^{-1}$, as well as at $\omega_1 \approx 0.73~\mathrm{s}^{-1}$ and $\omega_2 \approx 0.31~\mathrm{s}^{-1}$, corresponding to the two daughter waves. These frequency peaks satisfy the resonance condition $\omega_0 = \omega_1 + \omega_2$ to within 1\%, thus confirming the formation of a resonant wave triad; the spatial resonance condition, which is a well established experimental feature \cite{Joubaud2012,Bourget2013}, was also checked and confirmed. The relatively high forcing frequency results in steeply oriented constant phase lines for the forced wave field, which in turn yields strong and weak vertical and horizontal velocities, respectively. Thus, in figure \ref{fig:spectrogram} the frequency results for the $w$ velocity field more prominently feature the steeper primary wave, whereas the $u$ velocity field more readily highlights the shallower-propagating, subharmonic daughter waves. Notably, the $u$ velocity spectra also has a strong signal at $\omega =0$, which corresponds to a mean flow likely generated via entrainment by the horizontal phase velocity of the boundary forcing \citep{Bourget2013} (i.e. a mechanism resembling an Archimedean screw). One can also identify additional nonlinear interactions, albeit at significantly weaker amplitudes, between the primary and daughter waves at frequencies $\omega_0+\omega_1$ and $\omega_0+\omega_2$.

\begin{figure}
\centering
\includegraphics[width=1\columnwidth]{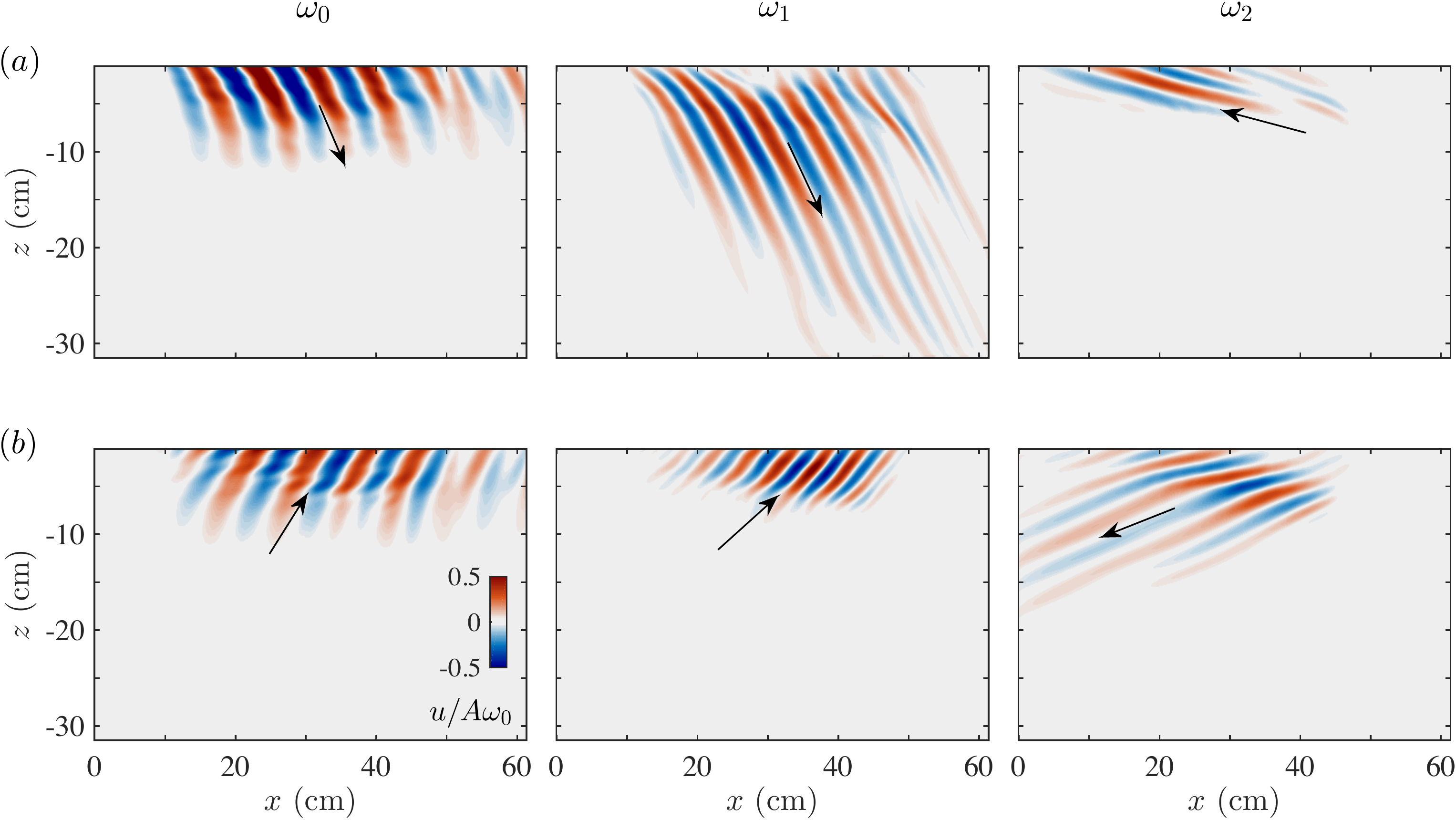}
\caption{Wave components comprising the wave field in the rightmost panel of figure \ref{fig:time evolution}. Panels in $(a)$ and $(b)$ correspond to the wave fields associated with the incident and reflected primary wave in the $N_1$ layer. Panels in the left column are the primary wave components oscillating at $\omega_0$; the middle and right columns correspond to the subharmonic daughter waves oscillating at $\omega_1$ and $\omega_2$, respectively. The orientation of the group velocity vectors are denoted by the black arrows in each plot. For clarity, we plot the $u$ velocity component normalized by the characteristic velocity scale $A \omega_0$.}
\label{fig:components}
\end{figure}

% Paragraph 10
Using the Hilbert transform filtering technique \citep{Mercier2008}, we decompose the experimental wave field into its different directionalities. Figure \ref{fig:components} presents the six wave components that comprise the wave field once PSI has been established (rightmost panel in figure \ref{fig:time evolution}, $t=30T_0$); for clarity, the $u$ velocity has been used to plot the constitutive waves. We see that the upper $N_1$ layer contains the incident primary wave and also a primary wave reflected from the relatively sharp $N_1$-to-$N_2$ transition in the stratification, as demonstrated by the results in the left-most panels of figure \ref{fig:components}$(a)$ and \ref{fig:components}$(b)$, respectively.  Each of these wave fields at the primary frequency $\omega_0$ spawn a pair of subharmonic daughter waves via PSI. Generally, a consequence of the spatial resonant triad condition is that the $\omega_1$-daughter wave will have a group velocity oriented in the same quadrant as that of the primary wave (i.e. $\mathrm{sgn}\left(k_1\right) = \mathrm{sgn} \left(k_0\right)$ and $\mathrm{sgn}\left(m_1\right) = \mathrm{sgn} \left(m_0\right)$) while the $\omega_2$-daughter wave has its group velocity oriented in the opposite direction (i.e. $\mathrm{sgn}\left(k_2\right) = -\mathrm{sgn} \left(k_0\right)$ and $\mathrm{sgn} \left(m_2\right) = -\mathrm{sgn} \left(m_0\right)$); as mentioned previously, we checked, and confirmed, that this spatial resonance condition was satisfied via now-standard procedures \citep{Joubaud2012}. Commensurate with these requirements,  the incident and reflected primary waves yield downward propagating $\omega_1$ and $\omega_2$ subharmonics, respectively, and since both subharmonic frequencies are smaller than $N_2$, in contrast to the primary wave, they are capable of penetrating into the lower stratification layer.

\begin{figure}
\centering
\includegraphics[width=0.7\columnwidth]{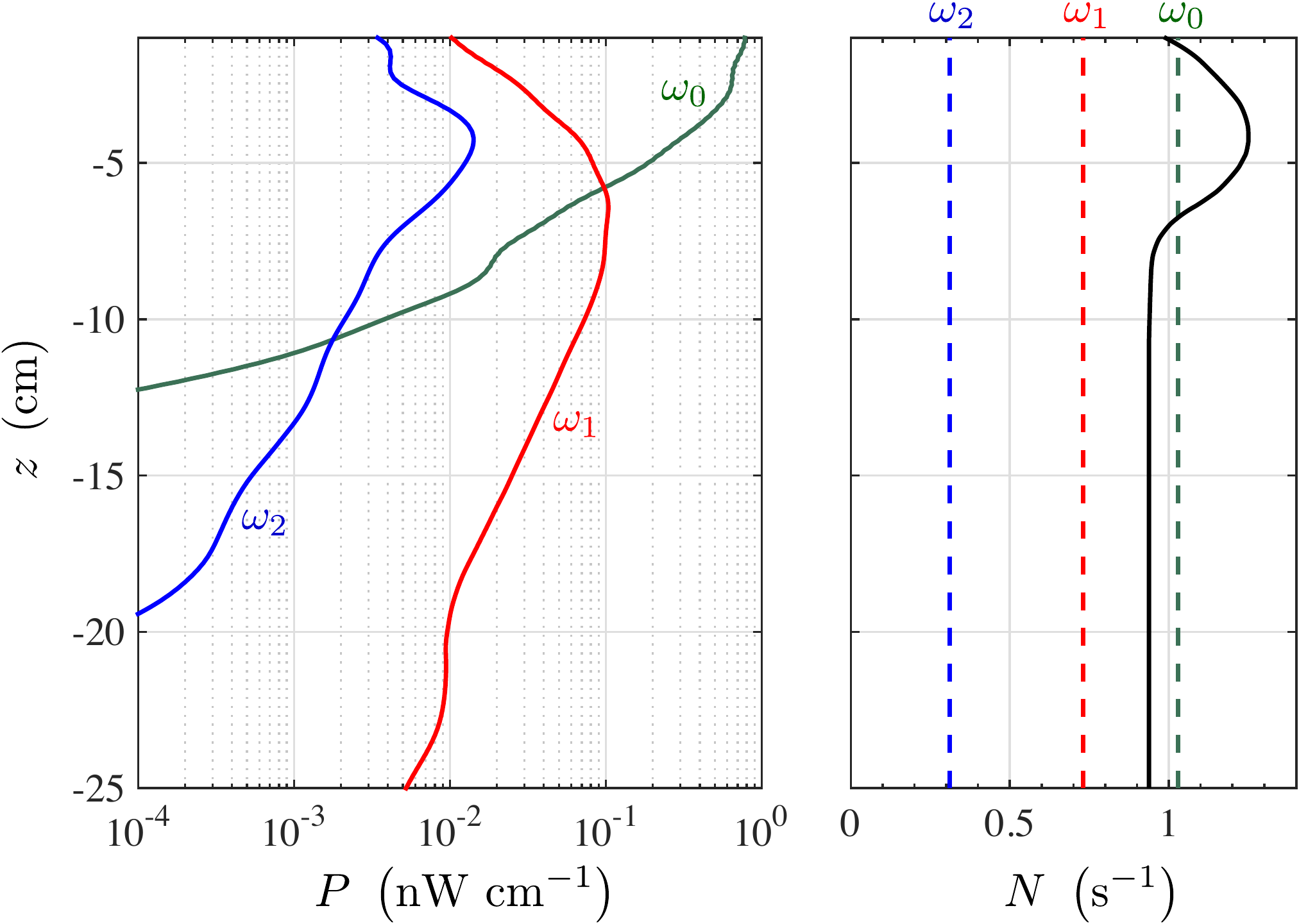}
\caption{(left) Power transmission profiles calculated by horizontally-integrating the vertical energy flux field for the downward wave components shown in figures \ref{fig:components}. (right) Buoyancy frequency profile (solid line) and corresponding temporal frequencies (dashed lines).}
\label{fig:power profiles}
\end{figure}

% Paragraph 11
From our PIV velocity field measurements, we calculate the energy flux field for the different frequency components using a recently established method \citep{Lee2014}. First, the stream function field $\psi(x,z)$ is obtained from the two-dimensional velocity field data via the relation $\left(u,w\right) = \left( -\partial_z \psi, \partial_x \psi \right)$ and path integration. The stream function field, as well as the background density and wave frequency, are subsequently incorporated into the expression for energy flux:
\begin{equation}
\boldsymbol{J} = \frac{\mathrm{i} \rho \psi}{\omega} \left( \left(\omega^2-N^2\right) w \, \boldsymbol{\hat{\underline{e}}}_x - \omega^2 u \, \boldsymbol{\hat{\underline{e}}}_z \right).
\end{equation}
With the energy flux field in hand, the vertical power radiation profiles (per unit width) are calculated via:
\begin{equation}
P = \int_L \langle \boldsymbol{J} \rangle \cdot \boldsymbol{\hat{\underline{e}}}_z \, \mathrm{d}x,
\end{equation}
where $L \in [0,60]$~cm is the horizontal domain of the experimental field of view, and the brackets $\langle \cdot \rangle$ denote time period averaging. We note that the primary wave field is weakly nonlinear by virtue of PSI taking place, and that the aforementioned definitions neglect viscosity. Nevertheless, we utilize this approach in order to obtain order of magnitude estimates of energy transmission.

% Paragraph 12
Figure \ref{fig:power profiles} presents the power transmission profiles computed by separately applying the energy flux calculations to each of the downward propagating wave components presented in figure \ref{fig:components}. As expected, there is no power transmission of relatively high frequency waves (i.e. $\omega>N_2$) into the weakly stratified lower layer. There is, however, notable power transmission for daughter waves at  frequency $\omega_1$ and some transmission at frequency $\omega_2$. More quantitatively, we find that $O(10\%)$ of the power of the primary wave is transmitted in the form of the daughter wave of frequency $\omega_1$, made possible by PSI, and $O(1\%)$ is transmitted at frequency $\omega_2$. The subsequent decay of this power with depth is attributable to viscous effects, which affect the daughter waves somewhat more than the primary wave, due to the smaller length scales of the former compared to the latter. We estimate the viscous dissipation of the vertical energy flux through the factor \citep{Lighthill1978}: 
\begin{equation}
\exp \left( \frac{-\nu k^3 N^4 z}{\omega^4 \sqrt{N^2-\omega^2}} \right).
\label{eq:viscous decay}
\end{equation}
Over the pycnocline, the primary wave vertical energy flux decreases by approximately $5\%$, while the vertical energy flux of the $\omega_1$- and $\omega_2$-daughter waves decrease by approximately $90\%$ and $70\%$, respectively, over the entire stratification. As can be deduced from \eqref{eq:viscous decay}, wave fields characterized by larger length scales experience less viscous decay, and therefore a reduced PSI threshold which allows an even larger fraction of the primary wave energy to penetrate the weakly stratified lower layer.

%%%%%%%%%%%%%%%%%%%%%%%%%%%%%%%%%%%%%%%%%%%%%%%%%%%%%%%%%
%%%%%%%%%%%%%%%%%%%%%%%%%%%%%%%%%%%%%%%%%%%%%%%%%%%%%%%%%

% Paragraph 13
In summary, our experiments provide the first demonstration of a novel process by which energy injected at relatively high frequencies into the strongly-stratified upper-layer of a water column can penetrate into an otherwise forbidden, weakly-stratified lower-layer lying beneath. The underlying mechanism is PSI, which extracts energy from the primary wave and distributes some of it into a pair of daughter waves, both of which are not evanescent, but propagating, in the lower layer. Two elements contribute to the development of PSI in our system: the broad-envelope of forcing \citep{Karimi2014,Bourget2014} and the near-total reflection of energy at the base of the pycnocline \citep{Zhou2013}. The consequent increase in the amplitude of the primary wave field induces faster growth rates of the instability. At higher Reynolds number and/or thinner pycnoclines, however, harmonic generation might diminish the PSI-driven transmission of wave energy into the lower layer \citep{Diamessis2014}. For the experimental configuration we used to demonstrate this scenario, we found the development of PSI to have an $O(10\%)$ effect with regards to power transmission, although more studies would be required to determine how this varies with the strength of the wave field, pycnocline thickness, and other system parameters such as $\omega_0$, $k_0$, $N_1$ and $N_2$. Building on this discovery, in future studies it would be of interest to investigate whether this scenario might have relevance to the ocean, which is also characterized by a strongly stratified upper layer and a weakly stratified abyss. Physical processes such as Langmuir circulation \citep{Thorpe2004,Polton2008} and  shear-driven instabilities \citep{Pham2009}, for example, are known to excite relatively high-frequency internal waves in the upper ocean, but relatively little is known about the magnitude and fate of the energy flux associated with them. In order for PSI to manifest in the ocean, the forcing mechanism would either have to be of large amplitude (since the growth rate of the instability is strongly dependent on the wave amplitude \citep{Bourget2013}), or persist for a sufficient number of forcing periods on a well established time scale. 

\begin{acknowledgements}
S.J.G. thanks the MIT-France Program for their support in establishing this collaboration. S.J., T.D. and P.O. thank ONLITUR (ANR-2011-BS04-006-01). T.P. acknowledges the support of the NSF (OCE-1357434). 
\end{acknowledgements}

%\bibliography{bibliographee}% Produces the bibliography via BibTeX.


\end{document}